\documentclass{PoS}
\usepackage[utf8]{inputenc}
\usepackage{graphicx}
\usepackage{bm}
\usepackage{color}
\usepackage{amsmath}
\usepackage{braket}
\usepackage{placeins}
\usepackage{multirow}
\usepackage{sidecap}
\title{Thermal modifications of quarkonia and heavy quark diffusion from a comparison of continuum-extrapolated lattice results to perturbative QCD}

\ShortTitle{Thermal modifications of quarkonia}

\author{\speaker{A.-L. Lorenz}\thanks{formerly Kruse.}, H. Sandmeyer, H.-T. Shu\\
        Fakultät für Physik, Universität Bielefeld, D-33615 Bielefeld, Germany\\
        E-mail: \email{alorenz@physik.uni-bielefeld.de}}

\author{ H.-T. Ding\\
Key Laboratory of Quark \& Lepton Physics (MOE) and Institute of Particle Physics,\\Central China Normal University, Wuhan 430079, China}

\author{ O. Kaczmarek\\
Key Laboratory of Quark \& Lepton Physics (MOE) and Institute of Particle Physics,\\Central China Normal University, Wuhan 430079, China\\
Fakultät für Physik, Universität Bielefeld, D-33615 Bielefeld, Germany
}
\author{H. Ohno\\
Center for Computational Sciences, University of Tsukuba, Tsukuba, Ibaraki 305-8577, Japan
}

\abstract{We investigate the in-medium modifications of heavy quarkonia in the vector channel and the heavy quark diffusion coefficient by comparing Euclidean correlators from the lattice to a perturbative spectral function. On the lattice side, we work with continuum extrapolated data from four different large and fine lattices with Clover-improved Wilson fermions in the quenched approximation at five temperatures (0.75, 1.1, 1.3, 1.5 and 2.25$T_c$). On the perturbative side, we use a combination of pNRQCD and vacuum asymptotics to describe the spectral function. After accounting for systematic errors, we obtain a spectral function that is suited to describe the bound state region. This spectral function describes charmonium well without a resonance peak at any of our analyzed temperatures above $T_c$, while we observe a thermally broadened resonance peak for bottomonium that is only melted at our largest temperature, $2.25T_c$. For the transport contribution we assume a Breit-Wigner shaped peak and find that the drag coefficient of charm quarks is larger than that of bottom quarks.}

\FullConference{37th International Symposium on Lattice Field Theory - Lattice2019\\
		16-22 June 2019\\
		Wuhan, China}

\newcommand{\dd}{\mathrm{d}}

\begin{document}

\section{Introduction}
To gain information on heavy quarkonium properties such as thermal modifications of bound states or transport coefficients, studies of the hadronic spectral function in the hot medium are needed \cite{ref-review,ref-rasmus}. On the lattice we cannot measure the spectral function directly, but instead calculate Euclidean correlators. In our previous work \cite{ref-ourpaper} we focused on pseudoscalar correlators. Now, we extend the method to the vector channel, where the correlation functions are given by
\begin{align}
G^{ii}(\tau)=\int_{\vec{x}}\braket{(\bar{\psi}\gamma_i \psi)(\tau,\vec{x}) (\bar{\psi}\gamma_i \psi)(0,\vec{0})}_c 
\end{align}
and related to the spectral function via
\begin{align}
\label{corr_kernel}
G^{ii}(\tau)=\int \limits_{0}^{\infty} \frac{\dd \omega}{\pi} \rho^{ii}(\omega)K(\omega,\tau)\quad \text{with kernel} \quad K(\omega,\tau)=\frac{\cosh(\omega(\tau-\frac{1}{2T}))}{\sinh(\frac{\omega}{2T})}.
\end{align}
Many methods for this ill-posed extraction problem have been proposed, for a recent review see \cite{ref-review}. In our approach, we compare continuum extrapolated correlators obtained from fine quenched lattices to a perturbative spectral function, constructed by combining vacuum asymptotics with pNRQCD calculations. Based on this we use a model that takes systematic uncertainties into account. For a similar study of the pseudoscalar channel see \cite{ref-ourpaper}, where we concluded that the corresponding ansatz in that channel describes the lattice data well. In the vector channel, we additionally have to include a term to describe the transport peak \cite{ref-transpeak}, that can not be expressed perturbatively. Therefore, we first fix the higher frequency spectral function by analyzing the difference of neighboring correlators and then further investigate the remaining contribution associated with transport effects.

\section{Perturbative Spectral Function} \label{sec-pertspf}
To obtain the perturbative spectral function, different expressions for different energy regimes need to be combined. At energies well above the threshold ($\omega \gg M$), the spectral function is given by vacuum asymptotics, e.g. in \cite{ref-uvasymptotics}. For energies around the threshold, \cite{ref-thresholdregion} applies pNRQCD with a real-time potential from hard thermal loop resummation. The two expressions are combined by introducing a matching factor $A^{match}$ that is chosen so that the transition from one regime to the other is steady and smooth at a matching point $\omega_{match}$. The details of this matching procedure are explained in \cite{ref-ourpaper}. The result describes the spectral function at intermediate to high energies, but overestimates it when approaching $\omega \approx 0$. Therefore, an exponential cut-off $\phi$ is introduced. With this, our perturbative spectral function reads
\begin{align}\label{eq-pertspf}
\rho_{pert}(\omega)=A^{match}\phi(\omega)\rho_{\text{NRQCD}}(\omega)\theta(\omega^{match}-\omega)+\rho_{vac}\theta(\omega-\omega^{match}).
\end{align}

\section{Lattice Setup}
\begin{figure}[htb]
\centering{
\includegraphics[width=0.45\textwidth]{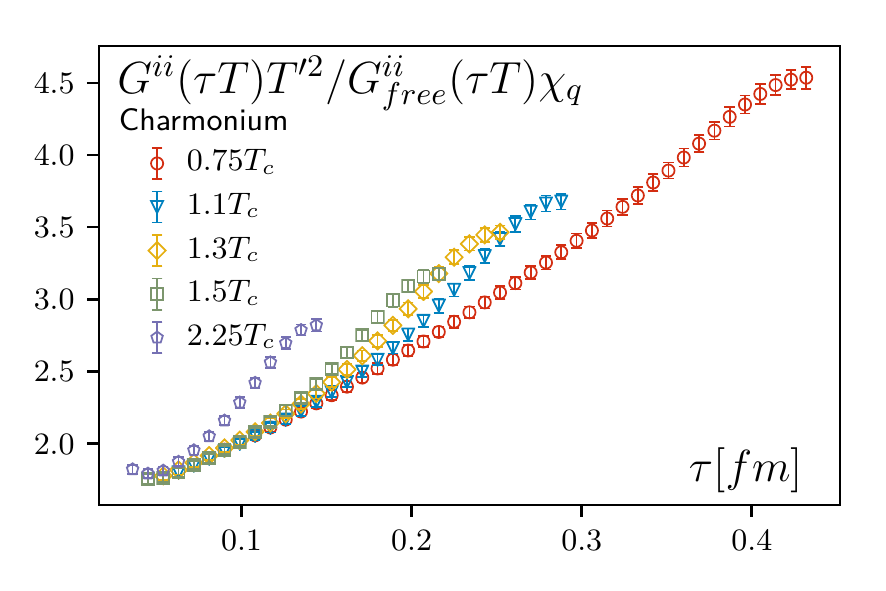}
\includegraphics[width=0.45\textwidth]{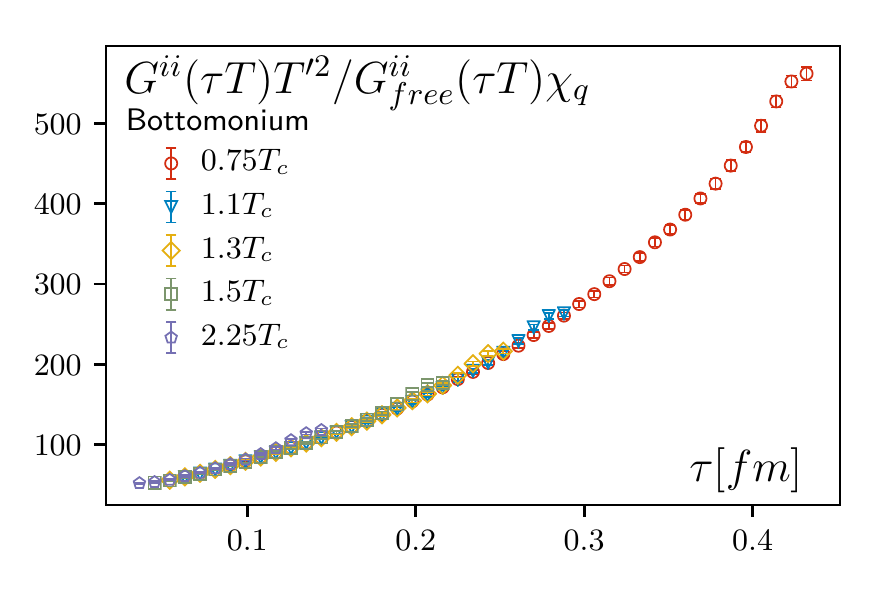}}
\caption{The continuum extrapolated correlators at different temperatures in the vector channel. Charmonium (\emph{left}) shows a stronger temperature dependence than bottomonium (\emph{right}), especially for larger $\tau$.\label{fig-corrsVV}}
\end{figure}

To compare our spectral function \eqref{eq-pertspf} to lattice results, we use continuum extrapolated lattice data coming from four different lattices generated in the quenched approximation with Wilson-clover valence quarks. An overview of the lattices is given in tab.~\ref{tab-latticesetup}. To obtain the continuum limit, we follow \cite{ref-contlim} and start with the renormalization. In the vector channel, we do not use renormalization constants and instead build a renormalization independent ratio with the quark number susceptibility $\chi_q T=G^{00}$. Only for extrapolating $\chi_q/T^2$ we have to use the renormalization constants. After this, we interpolate the correlators from the different lattices to the physical $J/\psi$ and $\Upsilon$ mass, before extrapolating to the continuum. With this, we obtain the correlators in fig.~\ref{fig-corrsVV}. Some qualitative conclusions can already be drawn at correlator level. For example, we can see that the temperature dependence is much larger for charmonium than for bottomonium. At small distances the correlators agree and then deviate with increasing $\tau$, meaning that there are more changes in the small frequency regime, most likely due to the growing transport contribution.
\begin{SCtable}
\centering
\scriptsize
\begin{tabular}{|c|c|c|c|c|c|c|}
\hline
$\beta$ & $r_0/a$ & $a$[fm]($a^{-1}$[GeV]) & $N_\sigma$ & $N_\tau$ & $T/T_c$ & $\#$ confs\\ \hline
\multirow{5}{*}{7.192}& & & & 48 & 0.75 & 237\\
   & & & & 32 & 1.1 & 476\\
    & 26.6 & 0.018(11.19) & 96 & 28 & 1.3 & 336\\
    & & & & 24 & 1.5 & 336\\
    & & & & 16 & 2.25 & 237\\ \hline
\multirow{4}{*}{7.394} &\multirow{4}{*}{33.8} & \multirow{4}{*}{0.014(14.24)}& \multirow{4}{*}{120} & 60 & 0.75 & 171\\
    & & & & 40 & 1.1 & 141\\
    & & & & 30 & 1.5 & 247\\
    & & & & 20 & 2.25 & 226\\ \hline
    & & & & 72 & 0.75 & 221\\
    & & & & 48 & 1.1 & 462\\
    7.544 & 40.4 & 0.012(17.01) & 144 & 42 & 1.3 & 660\\
    & & & & 36 & 1.5 & 288\\
    & & & & 24 & 2.25 & 237\\ \hline
    & & & & 96 & 0.75 & 224\\
    & & & & 64 & 1.1 & 291\\
    7.793 & 54.1 &  0.009(22.78) & 192 & 56 & 1.3 & 291\\
    & & & & 48 &1.5 & 348\\
    & & & & 32 & 2.25 & 235\\ \hline      
\end{tabular}
\caption{The four different $N_{\sigma}^3 \times N_\tau$ lattices used for the continuum extrapolation. The lattice spacing $a$ is determined using the Sommer scale (see \cite{ref-ourpaper}). On each lattice, correlators with six different $\kappa$-values \cite{ref-contlim} have been measured.} 
\label{tab-latticesetup}
\end{SCtable}
\section{Comparison}
Qualitatively, the perturbative correlator shows the same behaviour as the lattice data. To obtain quantitative results, we have to take uncertainties into account. We identified two main sources of systematic uncertainties. The first arises from the use of perturbative renormalization constants in the continuum extrapolation of $\chi_q/T^2$. We correct for this by introducing a normalization factor $A$. The second uncertainty lies in the definition of the pole mass in the threshold region of the perturbative calculation. We shift the spectral function by a constant $B$ to account for this. With these two corrections, we obtain a model spectral function for the large $\omega$ region $A\rho_{pert}(\omega -B)$. $A$ and $B$ are determined in a fit of the model correlator to the lattice data. To take the statistical errors into account, the whole procedure is conducted on bootstrap samples and the results are averaged. In \cite{ref-ourpaper} this method showed good agreement between perturbative calculations and the lattice correlators in the pseudoscalar channel. The fit describes the data almost perfectly, with only small changes to the perturbative spectral function, as $A$ is close to 1 and $B$ close to 0. For charmonium, we do not need an additional resonance peak to describe the data with the perturbative spectral function, while for bottomonium a thermally broadened resonance peak can persist up to about 1.5$T_c$.

\begin{figure}[!htb]
\centering{
\includegraphics[width=0.45\textwidth]{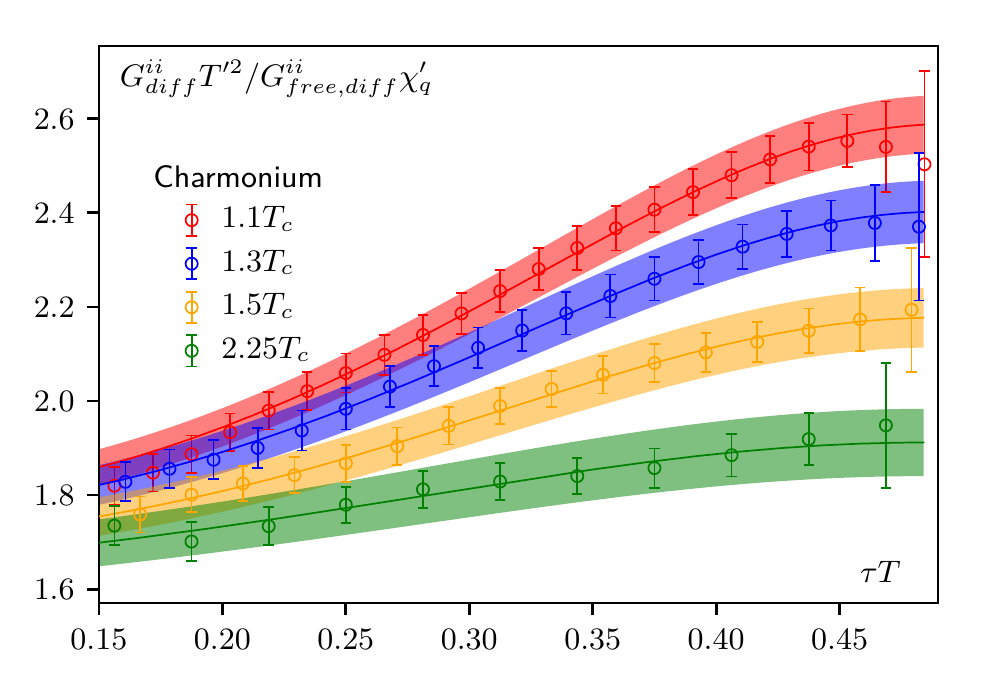}
\includegraphics[width=0.45\textwidth]{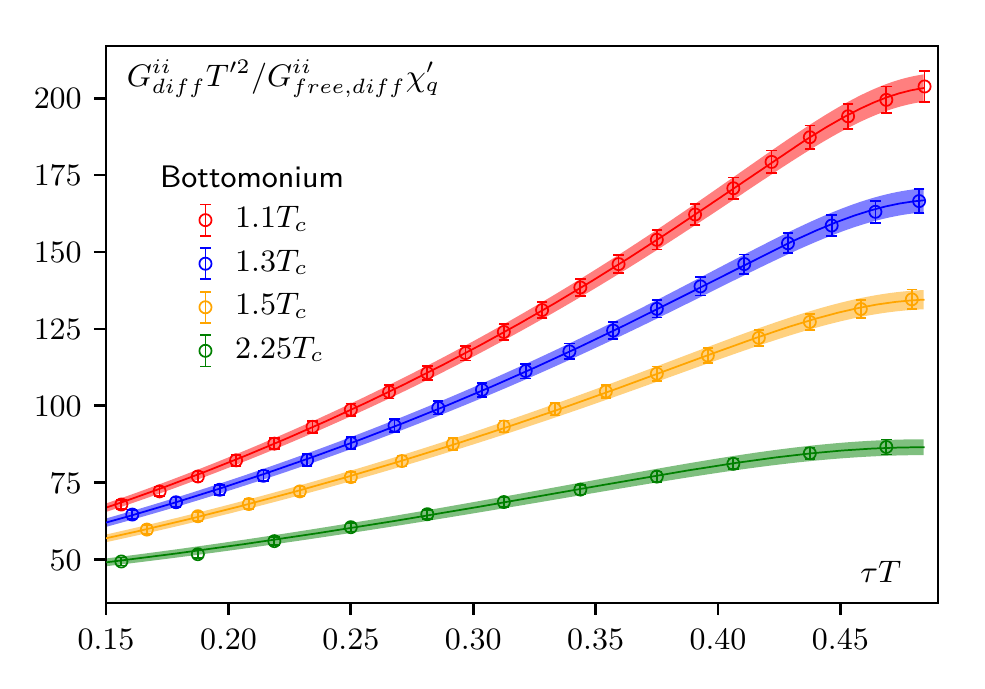}
\includegraphics[width=0.45\textwidth]{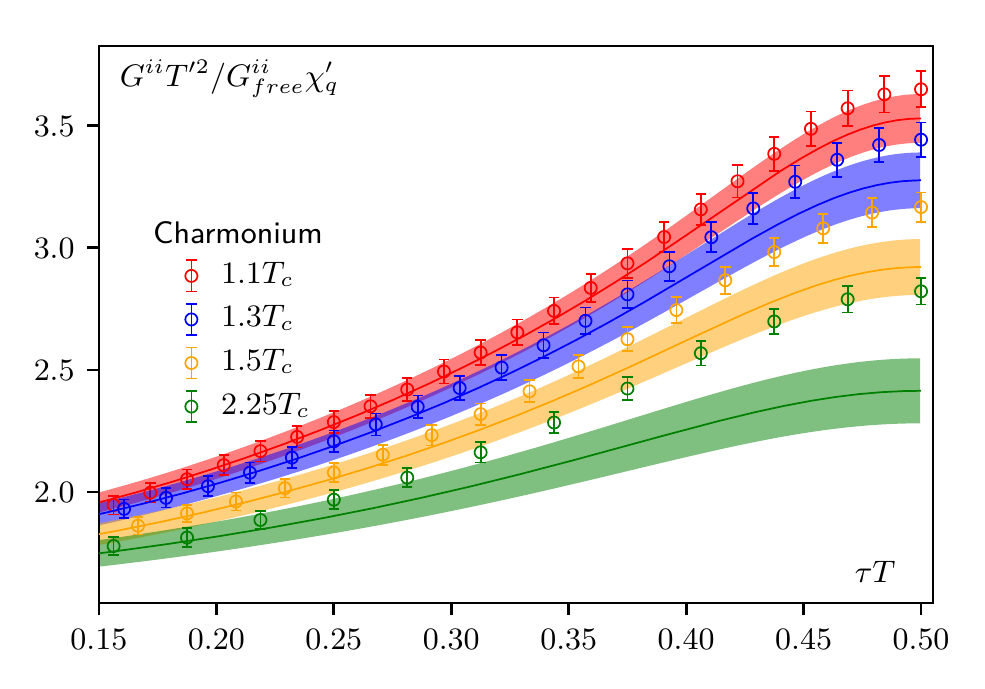}
\includegraphics[width=0.45\textwidth]{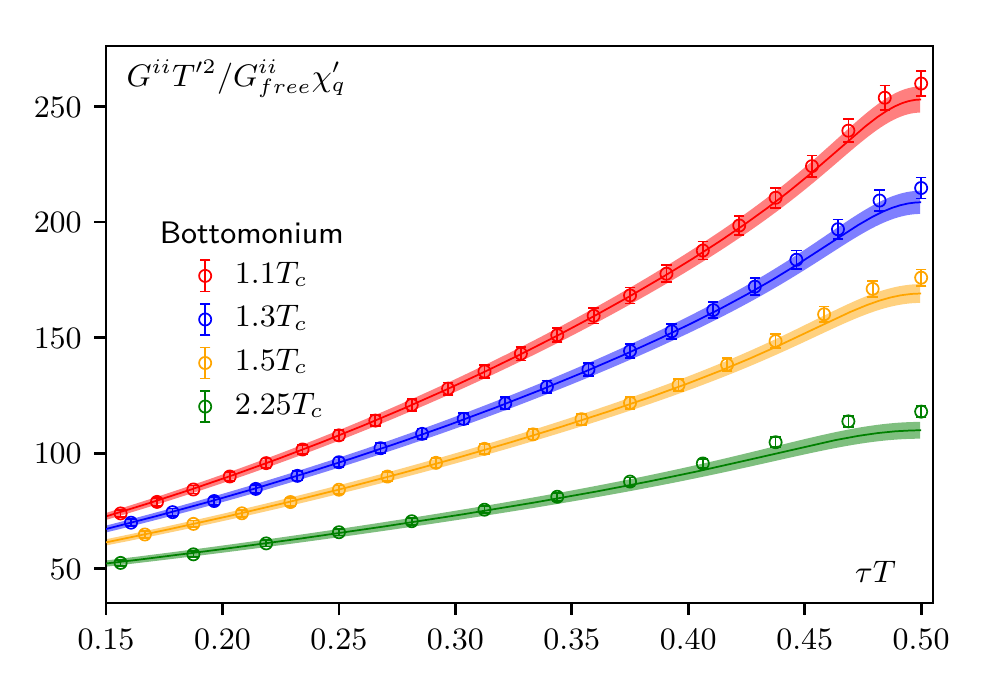}}
\caption{Fits to $G^{ii}_{diff}$ for charmonia (\emph{left}) and bottomonia (\emph{right}). The upper panel directly shows the fit result normalized with the free difference correlator. It can be seen that fit and data agree very well within errors. The lower panel shows the original correlator in comparison to the model correlator. The observed difference hints to a sizable transport contribution.}
	\label{fig-fitdiff}
\end{figure}

Because of heavy quark diffusion, a transport peak arises in the vector channel at small $\omega$. According to \cite{ref-transpeak} this transport peak can be modelled from Brownian motion and Langevin equations and has the shape of a Breit-Wigner peak
\begin{align}
\label{eq-breitwigner}
\rho_{trans}(\omega)=3D\chi_q \frac{\omega \eta_{D}^2}{\omega^2+\eta_{D}^2}\frac{1}{\cosh\left( \frac{\omega}{2\pi T} \right)},
\end{align}
where we additionally introduced a cut-off function for higher frequencies proposed in \cite{ref-breitwigner}. $D$ is the heavy quark diffusion coefficient and $\eta_D$ the drag coefficient. Due to this additional structure, the analysis of the vector channel is more challenging than that of the pseudoscalar channel. Since the perturbative spectral function does not include a description for the transport peak, we split our spectral function into two parts
\begin{align}
\rho_{model}(\omega)=\rho_{trans}(\omega)+A\rho_{pert}(\omega-B).
\end{align}
With this, we have four parameters ($A,B,2\pi TD,\eta_D$) that need to be determined. A direct fit fails due to the insensitivity of the correlator to the curvature introduced by $2\pi TD$. Instead we make use of the difference of neighboring correlators
\begin{align}
G^{ii}_{diff}(n_\tau)=G^{ii}(n_\tau)-G^{ii}(n_\tau +1)
\end{align}
that largely removes the transport contribution mostly present in the large $n_\tau$ region. We fit this using an ansatz consisting only of the perturbative part, i.e. $\rho^{model}=A\rho_{pert}(\omega-B)$. The obtained values of fit parameters A and B are listed in tab.~\ref{tab-Gdiff}. As seen from fig.~\ref{fig-fitdiff}, the fit results match the lattice data well. We also computed the correlation function $G^{ii}_{fit}$ from the fit spectral function $\rho_{model}$ and show the results as bands in the bottom two plots in Fig. 2. The differences between the bands and lattice data points hint to the transport contribution.

To further investigate the small $\omega$ part of the spectral function, we fix the values for $A$ and $B$ and calculate the difference of the lattice data and the fit results
\begin{align}
G^{ii}_{trans}(\tau T)=G^{ii}_{lat}-\int \frac{\text{d}\omega}{\pi} A\rho_{pert}(\omega-B)K(\omega,\tau).
\end{align}
Since the midpoint contains the largest contribution from transport effects and the curvature of the correlator is negligible, section \ref{sec-transport} focuses on $G^{ii}_{trans}(\tau T=0.5)$.

\begin{SCtable}
\centering
\scriptsize
\begin{tabular}{|c||c|c||c|c|}
\hline
 & \multicolumn{2}{c||}{Charmonium} & \multicolumn{2}{c|}{Bottomonium} \\ \hline
$T/T_c$ & $A$ &  $B/T$ & $A$ &  $B/T$\\ \hline
1.1 & 1.09(2) & 0.37(4) & 1.03(2) & 0.04(2) \\
1.3 & 1.07(2) & 0.16(5) & 1.01(1) & -0.05(2) \\
1.5 & 1.03(2) & 0.01(6) & 1.00(2) & -0.12(2) \\
2.25& 0.99(3) & -0.27(9) & 0.99(2) & -0.23(4)\\
\hline
\end{tabular}
\caption{Results for the high $\omega$ region from the fit of the perturbative spectral function to the lattice data $G^{ii}_{diff}$ in the vector channel.}
\label{tab-Gdiff}
\end{SCtable}

\subsection{Transport contribution}\label{sec-transport}

\begin{figure}[t]
	\includegraphics[width=0.45\textwidth]{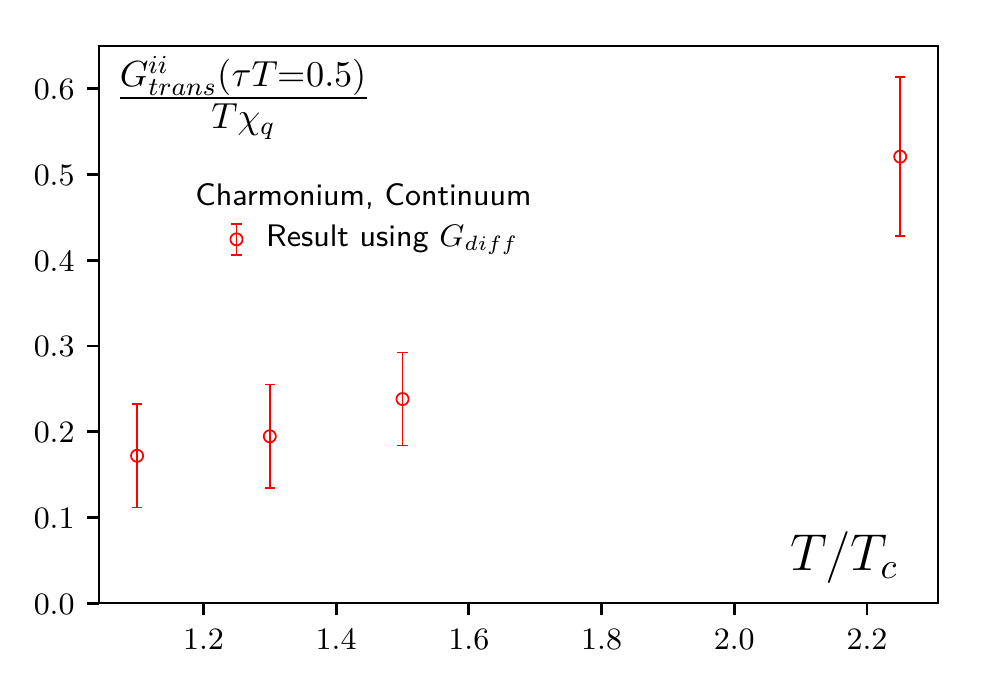}
	\includegraphics[width=0.45\textwidth]{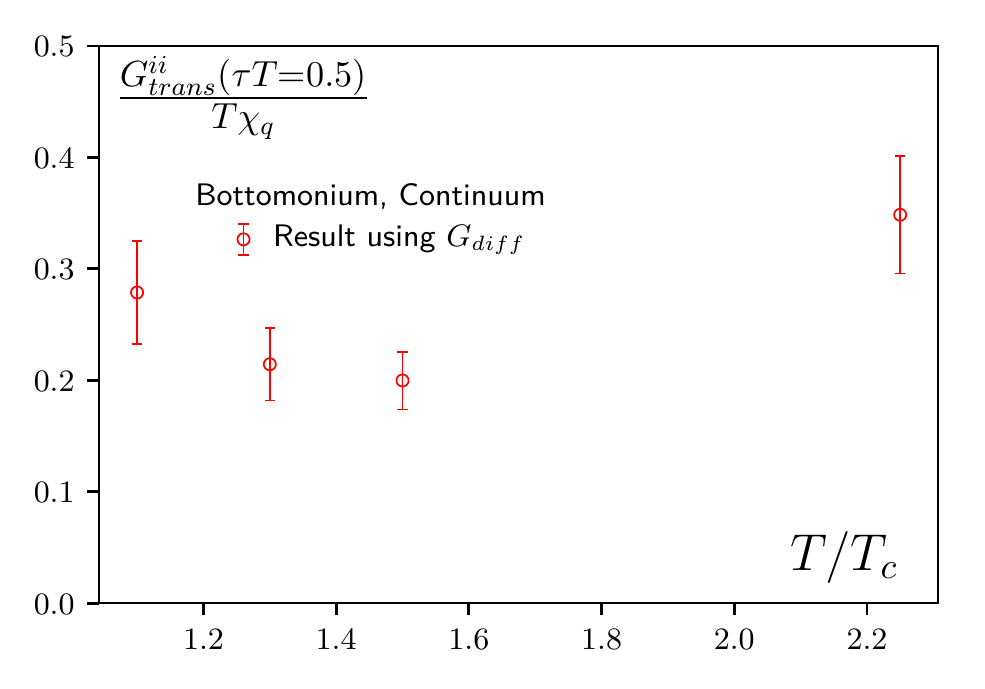}
	\caption{$G^{ii}_{trans}/(T\chi_q)$ for charmonium (\emph{left}) and for bottomonium (\emph{right}) at middle point as a function of temperature.}
	\label{fig:GtransMiddle}
\end{figure}

Fig.~\ref{fig:GtransMiddle} shows the temperature dependence of the transport contribution at the midpoint. It is clearly seen that the transport contribution for charmonium increases with temperature, while for bottomonium the results are almost constant in temperature within errors. Another observation is that for 1.5 and 2.25$T_c$ the charmonium result lies above the bottomonium result. This can be used to conclude a relation between the drag coefficients of charm and bottom quarks. To reach this goal, we expand the kernel and the cut-off term from \eqref{eq-breitwigner} at the midpoint around $\omega\approx 0$ as
\begin{align}
\frac{\cosh\left(\omega(1/2T-1/2T) \right)}{\sinh\left(\frac{\omega}{2T}\right)\cosh\left( \frac{\omega}{2\pi T} \right)}\approx& \frac{2T}{\omega}-\frac{(3+\pi^2)\omega}{12\pi^2 T}+\frac{(75+30+7\pi^4)\omega^3}{2880\pi^4 T^3}+\mathcal{O}(\omega^5)
\end{align}
We now insert \eqref{eq-breitwigner} into the correlator and make use of the above simplication. With the Einstein relation $\eta_D=T/MD$ we arrive at
\begin{align}
\frac{G^{ii}_{trans}(\tau T)}{\chi_q T}=\frac{T}{\pi M}\left( f_1 + f_2 + f_3  +\mathcal{O}(\omega^7)\right).
\end{align}
The dominant contribution comes from $f_1=2\tan^{-1}(\omega_{cut}/\eta_D)$ and higher orders can be neglected. When we insert the charmonium and bottomonium masses, we obtain
\begin{align}
\frac{G^{ii,c}_{trans}/\chi_{q}^c}{G^{ii,b}_{trans}/\chi_{q}^b}\approx \frac{M_b}{M_c}\frac{\tan^{-1}\left(\frac{\omega_{cut}}{\eta^{c}_D}\right)}{\tan^{-1}\left(\frac{\omega_{cut}}{\eta^{b}_D}\right)}.
\end{align}
The ratio of the masses is roughly 3 and when comparing this to the lattice data, we can conclude that $\tan^{-1}\left(\frac{\omega_{cut}}{\eta^{c}_D}\right)/\tan^{-1}\left(\frac{\omega_{cut}}{\eta^{b}_D}\right)<1$, which is only fulfilled when the charm quark drag coefficient $\eta_{D}^c$ is larger than that of the bottom quark $\eta_{D}^b$. We have checked the relative strength of drag coefficients using a Gaussian ansatz and $\eta^{c}_D > \eta^{b}_D$ also holds.

\subsection{Bound state region}\label{sec-boundstateregion}
\begin{figure}[htb]
\centering{
\includegraphics[width=0.43\textwidth]{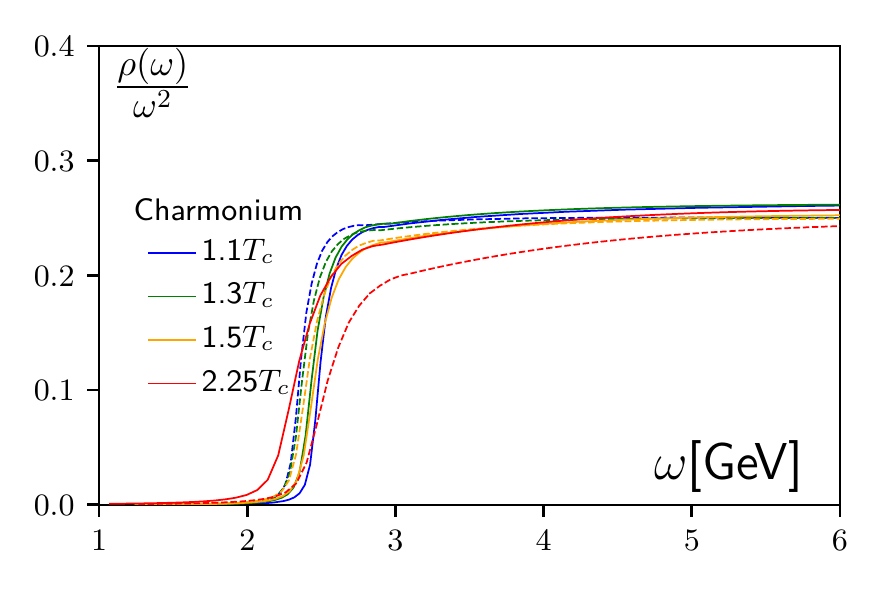}
\includegraphics[width=0.43\textwidth]{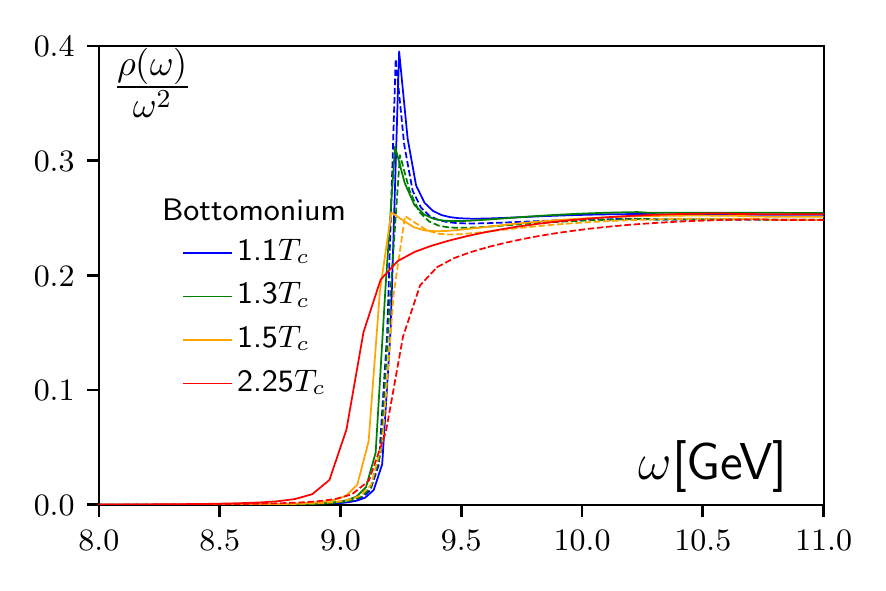}}
\caption{Spectral functions in the high $\omega$ region for charmonium and bottomonium in the vector channel. The dashed lines show the original perturbative spectral functions, while the solid lines show the modified spectral function from our fit to difference correlator. This figure does not include the transport peak.}
\label{fig-spfsVVboundstate}
\end{figure}

By fitting the differences of neighboring correlators we obtain a result for the high $\omega$ region of our spectral function in the vector channel. As in the pseudoscalar channel, the overall normalization $A$ is close to 1 and the mass shift is small. With these slight modifications, the perturbative spectral function is well-suited to describe the bound state region. The results are shown in fig.~\ref{fig-spfsVVboundstate}. For charmonium the model spectral function describes the data well without a resoncance peak while for bottomonium including one thermally broadened peak that vanishes at 2.25$T_c$ consistently describes the continuum extrapolated lattice data.
\section{Conclusion}
In this work, we found good agreement between our continuum extrapolated lattice correlators in the vector channel and a perturbative spectral function constructed by combining vacuum asymptotics with pNRQCD. By accounting for systematical uncertainties, we were able to present a modified version of the perturbative spectral function for the bound state regime that describes the correlator data almost perfectly. We split the analysis of the vector correlators in two seperate parts and fix the bound state region via fits to the difference of neighbouring correlators. We found that charmonium correlators in the vector channel can be well reproduced by perturbative spectral functions, where no resonance peaks for $J/\psi$ are needed at and above 1.1$T_c$, while for bottomonia correlators a thermally broadened resonance peak for $\Upsilon$ persists up to $\sim 1.5T_c$. From the analysis of the transport contribution we find that the charm quark drag coefficient is larger than the bottom quark drag coefficient. A further analysis of the transport peak we are currently following \cite{ref-wip} is to investigate the transport coefficients in more detail by comparing the lattice result at the midpoint to different ansätze for the transport peak.

\section{Acknowledgements}
This work was supported by the Deutsche Forschungsgemeinschaft (DFG, German Research Foundation) – project number 315477589 – TRR 211, NSFC under grant numbers 11775096 and 11535012.

\end{document}